\documentclass[aps,prl,twocolumn,superscriptaddress]{revtex4-1}

\usepackage{amsmath,amssymb}
\usepackage{graphicx}
\usepackage{epstopdf}
\usepackage{bm}
\usepackage{color}

\newcommand{\aRuCl}     {$\alpha$-RuCl$_3$}
\newcommand{\aRuIrCl}     {$\alpha$-Ru$_{1-x}$Ir$_x$Cl$_3$}
\newcommand{\cl}	{$^{35}$Cl}

\newcommand{\slr} 	{$T_1^{-1}$}
\newcommand{\slro} 	{$\mathcal{R}_0$}

\begin{document}

\title{Observation of a random singlet state in a diluted Kitaev honeycomb material}

\author{Seung-Ho Baek}
\email[]{sbaek.fu@gmail.com}
\affiliation{Department of Physics, Changwon National University, Changwon 51139, Korea}
\author{Hyeon Woo Yeo}
\affiliation{Department of Physics, Changwon National University, Changwon 51139, Korea}

\author{Seung-Hwan Do}
\affiliation{Department of Physics, Chung-Ang University,  Seoul 156-756,
Korea}
\author{Kwang-Yong Choi}
\affiliation{Department of Physics, Chung-Ang University,  Seoul 156-756,
Korea}
\author{Lukas Janssen}
\affiliation{Institut f\"ur Theoretische Physik and W\"urzburg-Dresden Cluster of Excellence ct.qmat, Technische Universit\"at Dresden, 01062 Dresden, Germany}
\author{Matthias Vojta}
\affiliation{Institut f\"ur Theoretische Physik and W\"urzburg-Dresden Cluster of Excellence ct.qmat, Technische Universit\"at Dresden, 01062 Dresden, Germany}
\author{Bernd B\"uchner}
\affiliation{Institut f\"ur Festk\"orper- und Materialphysik and W\"urzburg-Dresden Cluster of Excellence ct.qmat, Technische Universit\"at Dresden, 01062 Dresden, Germany}
\affiliation{IFW Dresden, Helmholtzstr. 20, 01069 Dresden, Germany}

\date{\today}


\begin{abstract}
We report a \cl\ nuclear magnetic resonance (NMR) study of the diluted Kitaev material \aRuIrCl\ ($x=0.1$ and $0.2$) where non-magnetic Ir$^{3+}$ dopants substitute Ru$^{3+}$ ions.
Upon dilution, the \cl\ spectra exhibit unusual large magnetic inhomogeneity, which sets in at temperatures below the Kitaev exchange energy scale. At the same time, the \cl\ spin-lattice relaxation rate \slr\ as a function of dilution and magnetic field unravels a critical doping of $x_c\approx 0.22$, towards which both the field-induced spin gap and the zero-field magnetic ordering are simultaneously suppressed, while novel gapless low-energy spin excitations dominate the relaxation process.
These NMR findings point to the stabilization of a random singlet phase in \aRuIrCl, arising from the interplay of dilution and exchange frustration in the quantum limit.
\end{abstract}


\maketitle


\section{Introduction}

The celebrated Kitaev honeycomb model \cite{kitaev06} features bond-dependent exchange frustration, leading to a quantum spin liquid (QSL) state with fractionalized Majorana-fermion excitations \cite{winter17,hermanns18,janssen19,takagi19}. To date, a number of candidate materials with dominant Kitaev interaction have been proposed, yet it remains a challenge to realize the genuine Kitaev QSL, as most of these materials display long-range magnetic order at low temperatures. Such order is induced by further symmetry-allowed interactions beyond the Kitaev model, whose interplay gives rise to a rich phase diagram depending on the relative strength of the Kitaev and non-Kitaev terms \cite{chaloupka10,chaloupka13,rau14,winter16,janssen17}.

Quenched disorder has a strong influence on spin correlations and thus offers an interesting route for achieving an exotic quantum state in Kitaev materials \cite{willans10,trousselet11,zschocke15,knolle19}.
In the pure Kitaev model, randomly placed vacancies have been shown to generate unscreened impurity moments \cite{willans10}, while random Kitaev couplings lead to a spin liquid with strongly enhanced low-energy excitations \cite{zschocke15,knolle19a}. More generally, strong exchange randomness may induce a random-singlet state, promoting liquid-like correlations against long-range magnetic order. In such a state, quantum spins form singlet dimers and resonating singlet clusters with a broad range of binding energies \cite{bhatt82,fisher94,singh10a,shiroka11,liu18a,kimchi18a,wu19,uematsu19,shiroka11}. For the disordered Heisenberg chain, the properties of the resulting state are essentially understood, leading to non-trivial scaling of thermodynamic quantities \cite{fisher94}. In higher dimensions, connections to valence-bond glasses have been discussed, together with the emergence of a minor fraction of orphan spins \cite{singh10a,kimchi18a}. However, little is known about the spin dynamics in such states.

In ruthenium trichloride \aRuCl, spin-orbit-coupled $J_\text{eff}=1/2$ moments of the Ru$^{3+}$ ions form a honeycomb lattice. Although the material displays magnetic long-range order at low temperatures, numerous experiments \cite{sandilands15,banerjee16,hirobe17,do17,kasahara18,wulferding20} have revealed unusual magnetic and thermodynamic properties suggesting that \aRuCl\ is proximate to a Kitaev QSL \cite{knolle14,banerjee16,nasu16}.
Our previous nuclear magnetic resonance (NMR) study in \aRuCl\ \cite{baek17} has shown that large magnetic fields induce a spin-gapped quantum paramagnet, subsequently confirmed and further explored by other experiments \cite{sears17,wolter17,wellm18,hentrich18,banerjee18}.
Spin dilution by non-magnetic impurities has been previously studied for \aRuCl, signaling a distinct disordered spin-liquid state where spin-glass features are weak or absent \cite{lampen-kelley17,do18,do20}. This may be contrasted to diluted honeycomb iridates, $A_2$Ir$_{1-x}$Ti$_x$O$_3$ ($A = \text{Na}, \text{Li}$), where a low-temperature spin glass has been found \cite{manni14,andrade14}.

In this paper, we investigate via \cl\ NMR the dilution effect on the field-induced spin-gapped phase and the dynamic spin excitations in \aRuCl. Our NMR results show that randomness induces strong and apparently gapless low-energy excitations. Over a range of dilutions and applied fields, these exist on top of a gapped paramagnetic background, while beyond $x_c\approx0.22$ the system displays spin-liquid-like features of a random-singlet state even at zero field.


\begin{figure*}
\centering
\includegraphics[width=0.95\linewidth]{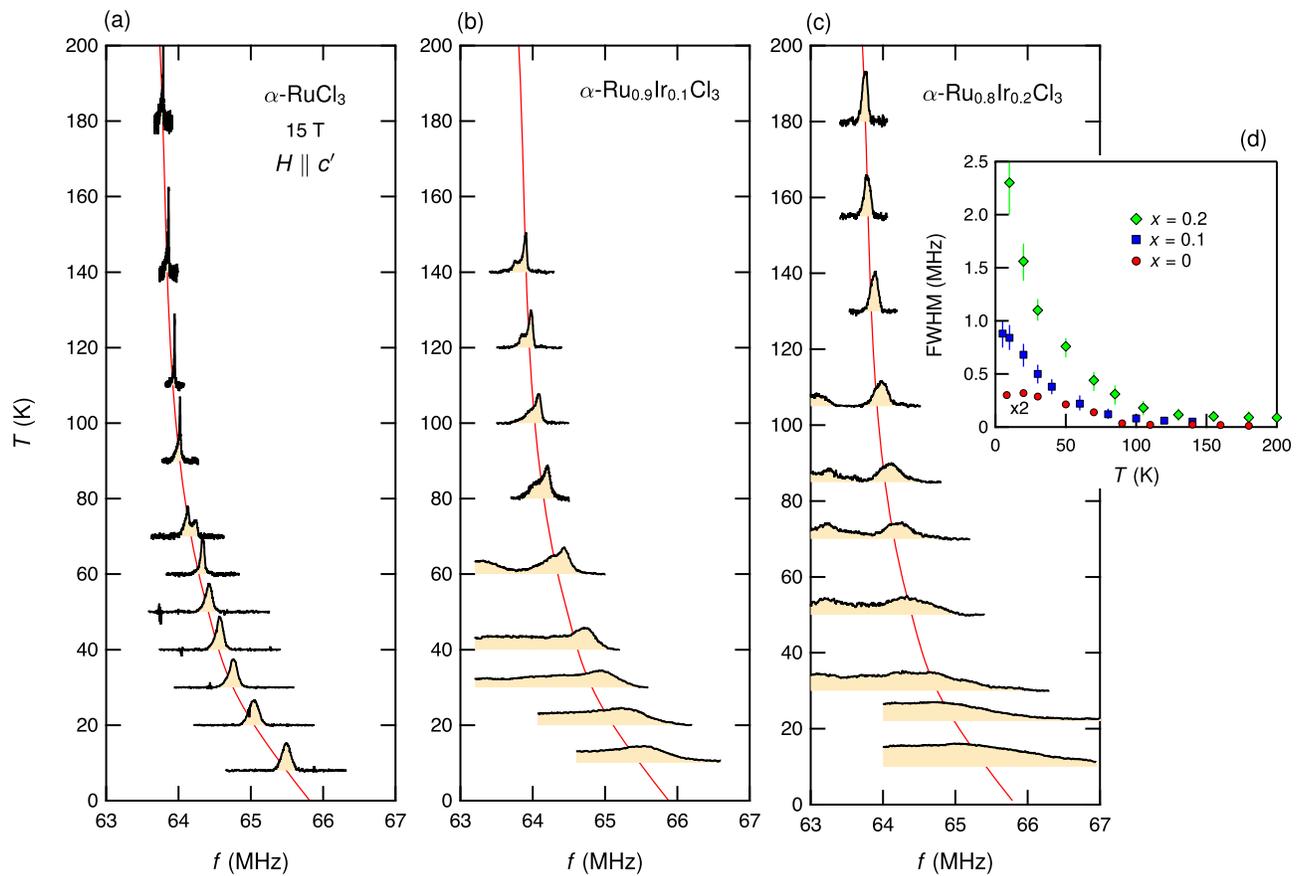}
\caption{
Temperature dependence of \cl\ NMR spectra at $15$\,T applied along the $c'$ axis (30$^\circ$ off the crystallographic $c$ axis) in \aRuIrCl: (a) $x=0$, (b) $x=0.1$, (c) $x=0.2$. Data for the undoped sample were taken from Ref.~\cite{baek17}. The \cl\ spectrum significantly broadens roughly below the Kitaev exchange scale $K$. Such a broadening becomes much stronger for $x=0.2$.
The red curves illustrate the average Knight shift (or average local magnetic susceptibility), which is essentially independent of $x$.
(d) FWHM of the main \cl\ line denoted by the solid curves as a function of temperature and Ir doping $x$. 
For the broad spectra at low temperatures, the FWHM was determined by taking and doubling the high-frequency part of the main line.
}
\label{spec}
\end{figure*}

\section{Experimental details}
\cl\ (nuclear spin $I=3/2$) NMR was carried out in \aRuIrCl\ single crystals ($x=0.1$ and 0.2) as a function of external field ($H$) and
temperature ($T$). The single crystals were synthesized as described in Ref.~\cite{do18}.
The samples were reoriented using a goniometer to achieve an accurate alignment along $\mathbf{H}$. The \cl\ NMR spectra were acquired by a standard spin-echo technique with a typical $\pi/2$ pulse length of 2--3\,$\mu$s. The nuclear spin-lattice relaxation rate \slr\ was obtained by fitting the recovery of the nuclear magnetization $M(t)$ after a saturating pulse to the fitting function
$1-M(t)/M(\infty)=A[0.9{\rm e}^{-(6t/T_1)^\beta}+0.1{\rm e}^{-(t/T_1)^\beta}]$, where $A$ is a fitting parameter and $\beta$ is the stretching exponent. 
Previously \cite{baek17}, we found that the three inequivalent \cl\ sites which form an extremely broad \cl\ spectrum owing to the strong quadrupole interaction mixed with the Zeeman interaction can be separated by tilting the crystal. We tracked the \cl\ line found at the highest frequency because it represents the minimal second order quadrupole interaction \cite{bennet}.  For $\sim 30^\circ$ off the crystallographic $c$ axis, which nearly  coincides with the angle between the octahedral edge of RuCl$_6$ and the $c$ axis, we observed a very narrow \cl\ line.   
All the NMR measurements in this work have been performed at the \cl\ line and the special direction of the external field is defined as the $c'$ axis.


\section{Results}
Figure~\ref{spec} shows \cl\ NMR spectra as a function of temperature for \aRuIrCl\ ($x=0$, $0.1$, $0.2$) at an external field of $15$\,T. 
At temperatures roughly above $100$\,K, the spectrum is weakly dependent on temperature and moderately broadens in proportion to $x$, which is reasonable based on the increase of chemical disorder. Below $\sim100$\,K, however, it rapidly broadens with lowering temperature and with increasing dilution.

For diluted samples, frequency-swept \cl\ spectra obtained for temperatures $\leq 60$\,K for $x=0.1$ and $\leq 105$\,K for $x=0.2$ reveal that another \cl\ line located at $\sim63$\,MHz is strongly quadrupole broadened, causing an extremely broad and nearly flat tail towards the low-frequency side. Fortunately, the main \cl\ line indicated by the red curve in Fig.~\ref{spec} remains identifiable.  Since the main line represents the \cl\ site in which the quadrupole interaction is minimal if not vanishing \cite{baek17}, we assume that its line broadening at low temperatures is of largely magnetic origin. The full width at half maximum (FWHM) of the main \cl\ line is presented in Fig.~\ref{spec}(d) as a function of dilution and temperature. For the broad spectra at low temperatures, the high-frequency part of the main line was taken and doubled. Note that the seeming divergence of the FWHM for $x=0.2$ is likely owing to the influence of a second \cl\ line.

\begin{figure*}
\centering
\includegraphics[width=0.9\linewidth]{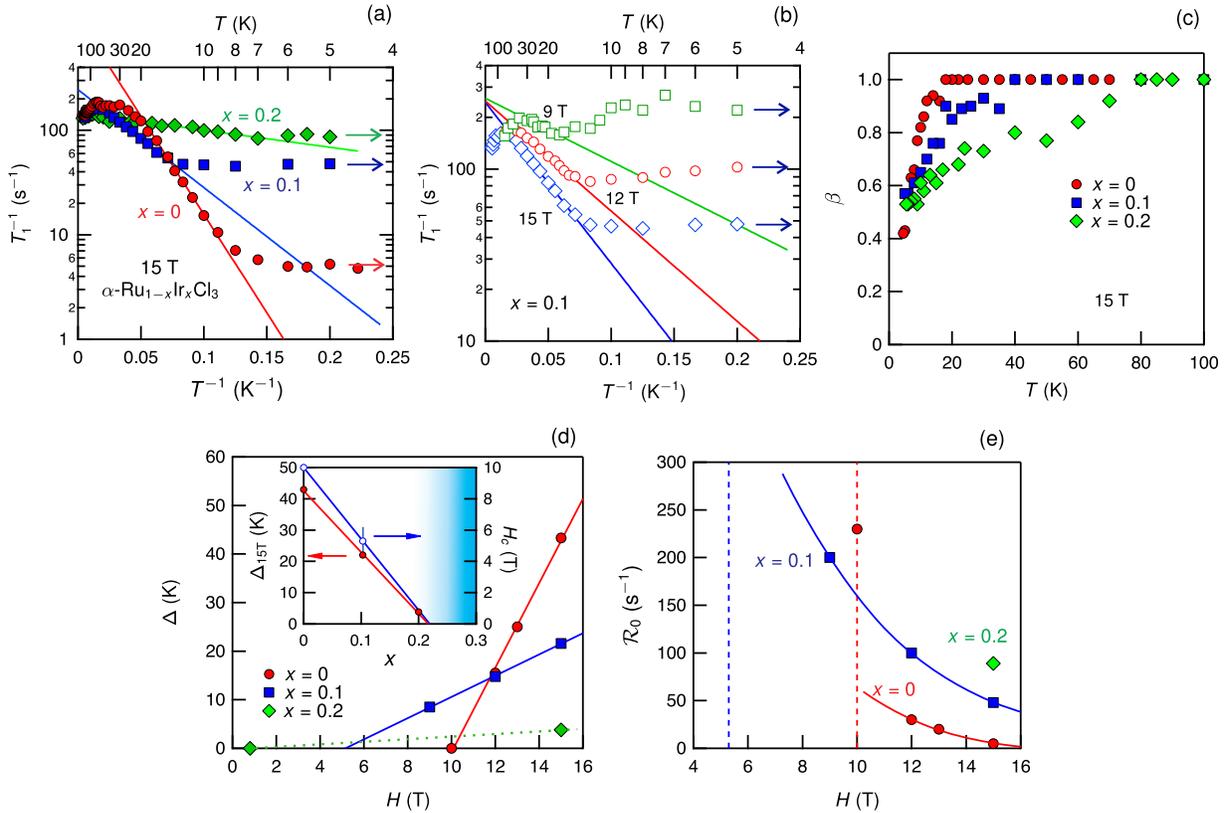}
\caption{Spin-lattice relaxation rate \slr\ vs.\ inverse temperature $1/T$: (a) at a fixed external field of $H=15$\,T for different Ir doping levels $x$, and (b) at $x=0.1$ for different $H$. The straight lines are Arrhenius fits to the chosen range of data that are linear in $1/T$ in the semi-log scale.
The spin gap $\Delta$ is suppressed with increasing $x$ and with decreasing $H$. In contrast, the plateau of \slr\ at low $T$, \slro, is strongly enhanced with increasing $x$ and with decreasing $H$.
(c) The stretching exponent $\beta$ as a function of $T$ and $x$ at 15 T. The temperature at which $\beta$ deviates from unity is rapidly enhanced with dilution, being consistent with the development of the strong magnetic inhomogeneity as shown in Fig. 1.
(d) $\Delta$ as a function of $H$ and $x$. With increasing $x$, the critical field $H_c$ below which $\Delta$ goes to zero decreases towards zero. The slope of $\Delta(H)$ are also suppressed  with increasing $x$, along with the decrease of $H_c$. Inset: The gap at 15\,T, $\Delta_\text{15\,T}$, (left axis) and the critical field, $H_c$, (right axis) go to zero simultaneously at $x_c\approx0.22$. $H_c$ for $x=0.2$ (the data point near 0.8\,T in the main panel) is estimated from the inset.
(e) \slro\ as a function of $H$ and $x$. Vertical lines denote the critical fields for a given $x$. ($H_c$ for $x=0.2$ is not shown.)  The diverging behavior of \slro\ towards $H_c(x)$ becomes stronger as $x$ increases. Data for the pristine sample were reproduced from Ref.~\cite{baek17}. Solid curves are guide to the eyes.
}
\label{t1t}
\end{figure*}

The development of strong magnetic inhomogeneity at low temperatures is unconventional, given that no spin-glass ordering has been detected down to $2$\,K in $\alpha$-Ru$_{0.8}$Ir$_{0.2}$Cl$_3$ via ac and dc susceptibility measurements \cite{do20}. Also, it is very unusual that with increasing dilution the temperature dependence of the main \cl\ line (Fig.~\ref{spec}), i.e., local spin susceptibility, remains unchanged, in contrast to the substantial line broadening. Furthermore, we note that the onset temperature of the broadening is very close to the Kitaev exchange energy scale $K\sim 90$\,K \cite{banerjee16} regardless of dilution. These features strongly suggest that the unusual magnetic inhomogeneity arises from the interplay between quenched disorder and the Kitaev interaction in the honeycomb lattice.

Having discussed the static aspects of the dilution effect in \aRuIrCl, we now turn to the evolution of low-energy spin dynamics with dilution, via the \cl\ spin-lattice relaxation rate \slr. To maintain consistency, \slr\ was measured on the peak of the main line denoted by the solid curves in Fig.~\ref{spec}, in magnetic fields applied along the $c'$ axis.
The results are presented in Fig.~\ref{t1t}(a-b) in an Arrhenius plot, i.e., \slr\ vs.\ inverse temperature on a semilog scale. In this plot, the straight line corresponds to an exponential decay, $T_1^{-1} \propto \exp(-\Delta/T)$, which yields the spin gap $\Delta$. Figure~\ref{t1t}(a) reveals that $\Delta$, defined in the temperature range $10$--$30$\,K, at a fixed field of $15$\,T is suppressed with increasing $x$, such that it almost vanishes at $x=0.2$. On top of this behavior, the field dependence of \slr\ at $x=0.1$, Fig.~\ref{t1t}(b), indicates that $\Delta$ decreases with decreasing $H$ at a given $x$, as observed in the pristine sample \cite{baek17}. Figure 3(c) shows the stretching exponent $\beta$ in \slr, which is a measure of spatial inhomogeneity. Clearly, the onset temperature in which $\beta$ deviates from unity is progressively enhanced with increasing dilution, which is consistent with the development of strong magnetic inhomogeneity with dilution (see Fig. 1).

The field and doping dependence of the spin gap $\Delta$ is summarized in Fig.~\ref{t1t}(d). Clearly, the critical field $H_c$, above which the spin-gapped quantum paramagnet emerges out of long-range magnetic order \cite{baek17}, is strongly reduced for $x=0.1$. As shown in the inset of Fig.~\ref{t1t}(d), a linear extrapolation of $H_c$ enables us to define the critical dilution $x_c\approx0.22$. This is consistent with previous results which show that zero-field order is suppressed for $x>x_c$ \cite{lampen-kelley17,do18}.
Figure~\ref{t1t}(d) also reveals that the slope of $\Delta$ vs.\ $H$ is rapidly reduced with increasing $x$. This means that the gapped quantum paramagnet itself is suppressed with dilution in the entire field range measured. Even more remarkably, the spin gap extracted at $15$\,T, $\Delta_\text{15T}$, also extrapolates to zero at $x_c\approx0.22$, precisely as does $H_c$, as shown in the inset of Fig.~\ref{t1t}(d). Together, this indicates a quantum transition into a gapless disordered state at $x_c$ existing over a large range of applied fields.

As also seen in Fig.~\ref{t1t}(a,b), the nominally gapped state at elevated fields is highly non-trivial: \slr\ flattens out toward a constant value at low temperatures, which is strongly enhanced with increasing $x$ and with decreasing $H$ (horizontal arrows). We ascribe the \slr\ plateau to abundant in-gap spin excitations, and identify it with the low-energy scale of the spin dynamics, \slro.
\slro\ as a function of $x$ and  $H$ is shown in Fig.~\ref{t1t}(e).
While \slro\ for $x=0$ is very small at 15\,T, it increases rapidly with decreasing $H$ and abruptly jumps to a large relaxation value at around $H_c=10$\,T. It should be noted that \slro\ preserves its meaning only when $\Delta$ is finite, and thus the discontinuous change of \slro\ at $H_c$ implies an alteration of its underlying mechanism. For $x=0.1$, \slro\ shows a strong increase with decreasing $H$, diverging toward the critical field $H_c\approx 5.2$\,T.  With further dilution to $x=0.2$, the gap behavior is negligibly weak even at $15$\,T, and thus it is difficult to measure the field dependence of the gap.


\section{Discussion}
Taken together, magnetic dilution in \aRuCl\ has a twofold effect: It creates apparently gapless spin excitations which progressively fill in the field-induced spin gap, as evidenced by our NMR data, and it suppresses zero-field magnetic order \cite{lampen-kelley17}. Beyond a critical doping level $x_c\approx0.22$, this results in a strongly inhomogeneous paramagnetic state which features distinct spin-liquid properties.

The properties of this disorder-stabilized spin liquid appear compatible with a random-singlet state: A collection of spin singlets with broadly distributed binding energies down to zero implies gapless and scale-invariant relaxation behavior, as seen in Fig.~\ref{t1t}(a) at $x=0.2$. It also implies that the relaxation remains gapless over a significant range of magnetic fields, as the applied field polarizes singlets with small binding energy, but leaves those with large binding energy intact. This also results in a field-driven reduction of the low-temperature relaxation rate, as seen in Fig.~\ref{t1t}(b). Moreover, a random-singlet state yields strongly-broadened NMR lines and the spatial distribution of \slr\ [Fig. 2(c)], as the applied field converts the broad range of binding energies into a broad distribution of local magnetizations. Lastly, a random-singlet picture is also supported by the power law and scaling behavior of the specific heat and the magnetic susceptibility observed in $\alpha$-Ru$_{0.8}$Ir$_{0.2}$Cl$_3$ \cite{do20}, although the power-law specific heat is in principle equally consistent with a bond-disordered Kitaev model \cite{willans10, knolle19a}. We note that the critical doping level is not very far from the percolation threshold $x_p \approx 0.303$~\cite{jacobsen14}. Assuming dominant nearest-neighbor exchange, the spin-liquid phase is then characterized by a large number of finite-size clusters that are only weakly coupled and contribute nontrivially to the observed spin dynamics.

We further note that even nominally clean high-quality \aRuCl\ displays a spin relaxation which, at $15$\,T, is disorder dominated below $6$\,K, Fig.~\ref{t1t}(a) \cite{baek17}. Similarly, a notable line broadening is visible below $80$\,K, Fig.~\ref{spec}(d). These features likely arise from residual point defects (or stacking faults) and indicate a remarkably strong sensitivity of the quantum paramagnet to quenched disorder.

Microscopically, the emergence of a random-singlet state in diluted \aRuCl\ must be driven by the interplay of quenched disorder, strong Kitaev and $\Gamma$ couplings \cite{winter16,janssen17}, and strong quantum effects. In contrast, a semiclassical treatment of the relevant models with quenched disorder results in strong spin-glass behavior \cite{andrade14}, which is not seen experimentally. We also note that dilution of a pure Kitaev model is unlikely consistent with the data, as the vacancy-induced degrees of freedom will be polarized and thus quenched by moderate fields at least for weak dilution \cite{willans10}, as opposed to the relaxation behavior seen for nominally clean \aRuCl. We consider it likely that a local substitution $\text{Ru} \rightarrow \text{Ir}$ not only removes a spin, but also substantially modifies the exchange couplings near the dopant site. Studying the combined effect of site and bond disorder in quantum Heisenberg-Kitaev-$\Gamma$ models is an important task for the future.


\section{Summary}
Our NMR measurements have demonstrated that site dilution of the Kitaev material \aRuCl\ results in a gapless spin liquid beyond a critical doping level $x_c\approx0.22$. The collected data appear compatible with a random-singlet state, which is demonstrated here for the first time in a Kitaev material. Our study calls for more detailed investigations of other types of disorder in \aRuCl\ as well as for a detailed microscopic understanding of disordered Heisenberg-Kitaev-$\Gamma$ models.


\begin{acknowledgments}
We thank E. C. Andrade, J. van den Brink, and A. U. B. Wolter for discussions and collaborations on related work.
This work was supported by the National Research Foundation of Korea (NRF) grant funded by the Korea government(MSIT) (NRF-2020R1A2C1003817) as well as by the Deutsche Forschungsgemeinschaft (DFG) through SFB 1143 (project id 247310070), the W\"urzburg-Dresden Cluster of Excellence on Complexity and Topology in Quantum Matter -- \textit{ct.qmat} (EXC 2147, project id 390858490), and the Emmy Noether program (JA 2306/4-1, project id 411750675).
\end{acknowledgments}


\bibliography{mybib}

\end{document}